\documentclass{elsart}

\usepackage{times}
\usepackage{helvetic}

\usepackage{graphicx}

\def\[{\begin{equation}}
\def\]{\end{equation}}

\begin{document}
\begin{frontmatter}

\title{Spatial competition and price formation}

\author[eth]{Kai Nagel,\thanksref{kn}}
\author[cowles]{Martin Shubik,\thanksref{ms}}
\author[nbi]{Maya Paczuski,\thanksref{mp}}
\author[nbi]{Per Bak\thanksref{pb}}

\address[eth]{Dept.\ of Computer Science, ETH Z\"urich, Switzerland}
\address[cowles]{Cowles Foundation for Research in Economics, 
Yale University,
New Haven, Connecticut}
\address[nbi]{Niels Bohr Institute, University of Copenhagen, Denmark}

\thanks[kn]{nagel@inf.ethz.ch. 
ETH Zentrum IFW B27.1, 8092~Z{\"u}rich, Switzerland}
\thanks[ms]{martin.shubik@yale.edu}
\thanks[mp]{paczuski@alf.nbi.dk}
\thanks[pb]{bak@alf.nbi.dk}

\maketitle

\begin{abstract}
  We look at price formation in a retail setting, that is, companies
  set prices, and consumers either accept prices or go someplace else.
  In contrast to most other models in this context, we use a
  two-dimensional spatial structure for information transmission, that
  is, consumers can only learn from nearest neighbors.  Many aspects
  of this can be understood in terms of generalized evolutionary
  dynamics. In consequence, we first look at spatial competition and
  cluster formation without price.  This leads to establishement size
  distributions, which we compare to reality.  After some theoretical
  considerations, which at least heuristically explain our simulation
  results, we finally return to price formation, where we demonstrate
  that our simple model with nearly no organized planning or
  rationality on the part of any of the agents indeed leads to an
  economically plausible price.
\end{abstract}

\begin{keyword}
econophysics; economics; simulation; markets
\end{keyword}

\end{frontmatter}
%%%%%%%%%%%%%%%%%%%%%%%%%%%%%%%%%%%%%%%%%%%%

\section{Introduction}

There are several basic concepts which lie at the heart of economic
theory.  They are the "economic atom" which is usually considered to
be the individual, profits, money, price and markets and the more
complex organism the firm.  Much of economic theory is based on
utility maximizing individuals and profit maximizing firms.  The
concept of a utility function attributes to individuals a considerable
amount of sophistication.  The proof of its existence poses many
difficult problems in observation and measurement.  In this study of
market and price formation we consider simplistic social individuals
who must buy to eat and who look for where to shop for the best price.
In this foray into dynamics we opt for a simple model of consumer
price formation. Our firms are concerned with survival rather than a
sophisticated profit maximization.  Yet we relate these simple
behaviors to the more conventional and complex ones.

A natural way to approach the economic physics of monopolistic
competition is to introduce space explicitly.  For much of economic
analysis of competition space and information are critical factors.
The basic aspects of markets involve an intermix of factors, such as
transportation costs and delivery times which depend explicitly on
physical space.  But for pure information, physical distance is less
important than direct connection.  For questions concerning the growth
of market areas, the spatial representation is appropriate.
Consideration of space is sufficient to provide a justification of
Chamberlin's model of monopolistic competition as is evident from the
work of Hotelling~\cite{Hotelling}. Furthermore it is reasonably
natural to consider space on a grid with some form of minimal
distance.  Many of the instabilities found in economic models such as
the Bertrand model are not present with an appropriate grid.

When investigating these topics, one quickly finds that many aspects
of price formation can be understood in terms of generalized
evolutionary dynamics.  In consequence, our first models in this paper
study spatial competition and cluster formation without the generation
of price (Sec.~\ref{sec:spatial}).  This generates cluster size
distributions, which can be compared to real world data.  We spend
some time investigating theoretical models which can explain our
simulation data (Sec.~\ref{sec:theo}).  We then, finally, move on to price
formation, where we implement the price dynamics ``on top'' of the
already analyzed spatial competition models (Sec.~\ref{sec:prices}).
The paper is concluded by a discussion and a summary.

%%%%%%%%%%%%%%%%%%%%%%%%%%%%%%%%%%%%%%%%%%%%
\section{Related work}

The model is an open one related to the partial equilibrium models of
much of micro-economics.  In particular money and its acceptance in
trade is taken as a primitive concept.  There is a literature on the
acceptance of money both in a static equilibrium context (see for
example~\cite{Kiyotaki:Wright:money}) and in a "bootstrap" or dynamic
context (see for example \cite{Bak:etc:money,Sneppen:money}).  These
are extremely simple closed models of the economy where each
individual is both a buyer and seller.  Eventually we would like to
construct a reasonable model where the acceptance of money, the
emergence of competitive price and the emergence of market structure
all arise from the system dynamics.  This will call for an appropriate
combination of the features of the model presented here with the
closed models noted above. We do not pursue this further here.
Instead by taking the acceptance of money as given our observations
are confined to the emergence of markets and the nature of price. The
static economic theories of monopoly and mass homogeneous competitive
equilibrium provide natural upper and lower benchmarks to gauge market
behavior.  The intermediate zone between $n=2$ and very many is
covered in the economic literature by various oligopoly models, of
which those of Cournot~\cite{Cournot:book}, Bertrand~\cite{Bertrand}
and Chamberlin~\cite{Chamberlin:book} serve as exemplars.  The
Chamberlin model unlike the earlier models stresses that all firms
trade in differentiated goods. They are all in part differentiated or
partially monopolistic.  When one considers both information and
physical location this is a considerable step towards greater realism.
%
%\yyyy{Martin: Unfortunately I don't understand the second-last
%  sentence (who is ``they''); and (in consequence?) I don't see the
%  connection to the last sentence.}
%
Other work on evolutionary or behavioral learning in price formation
are Refs.~\cite{Hehenkamp:etc:note,Hehenkamp:sluggish,Brenner:prices}.

%%  Farmer --- $\Delta p(t) \propto D(t) - S(t)$.
%%  
%%  Bak, Norrelyke, Shubik --- individual maximization from calculus.
%%  Adaptive version should give same result.
%%  
%%  Takayasu --- individual reservation prices.  Not coupled to anything.
%%  (Obwohl man nat"urlich argumentieren k"onnte, da"s diejenigen, die mit
%%  den (reservation) Preisen immer weiter runter gehen, die producers
%%  sind, denen das Geld ausgeht, und diejenigen, die mit den reservation
%%  prices immer weiter rauf gehen, die consumers sind, denen das food
%%  ausgeht.) 
%%  
%%  Bak, Paczuski, Shubik --- individual reservation prices.  Not coupled
%%  to anything.  (Unbiased version Takayasu? -- non-behavioral limiting
%%  case?)

%\processdelayedfloats
%%%%%%%%%%%%%%%%%%%%%%%%%%%%%%%%%%%%%%%%%%%%%%%%%%%%
\section{Spatial competition}
\label{sec:spatial}

As mentioned in the introduction, we will start with spatial models
without price.  We will add price dynamics later.

\subsection{Basic spatial model (domain coarsening)}
\label{sec:basic}

We use a 2-dimensional $N = L \times L$ grid with periodic boundary
conditions.  Sites are numbered $i=1..N$.  Each site belongs to a
cluster, denoted by $c(i)$.  Initially, each site belongs to
``itself'', that is, $c(i) = i$, and thus cluster numbers also go from
$1$ to $N$.

The dynamics is such that in each time step we randomly pick a
cluster, delete it, and the corresponding sites are taken over by
neighboring clusters.  Since the details, in particular with respect
to the time scaling, make a difference, we give a more technical
version of the model.  In each time step, we first select a cluster
for deletion by randomly picking a number $C$ between $1$ and $N$.
All sites belonging to the cluster (i.e.\ $c(i) = C$) are marked as
``dead''.  We then let adjoining clusters grow into the ``dead'' area.
Because of the interpretation later in the paper, in our model the
``dead'' sites play the active role.  In parallel, they all pick
randomly one of their four nearest neighbors.  If that neighbor is not
dead (i.e.\ belongs to a cluster), then the previously dead site will
join that cluster.  This step is repeated over and over, until no dead
sites are left.  Only then, time is advanced and the next cluster is
selected for deletion.

\begin{figure}[t]
  \begin{center}
%    \ncludegraphics[height=0.4\textheight]{basic-1-gz.eps}
%    \ncludegraphics[height=0.4\textheight]{basic-2-gz.eps}
\centerline{
    \includegraphics[width=0.4\textwidth]{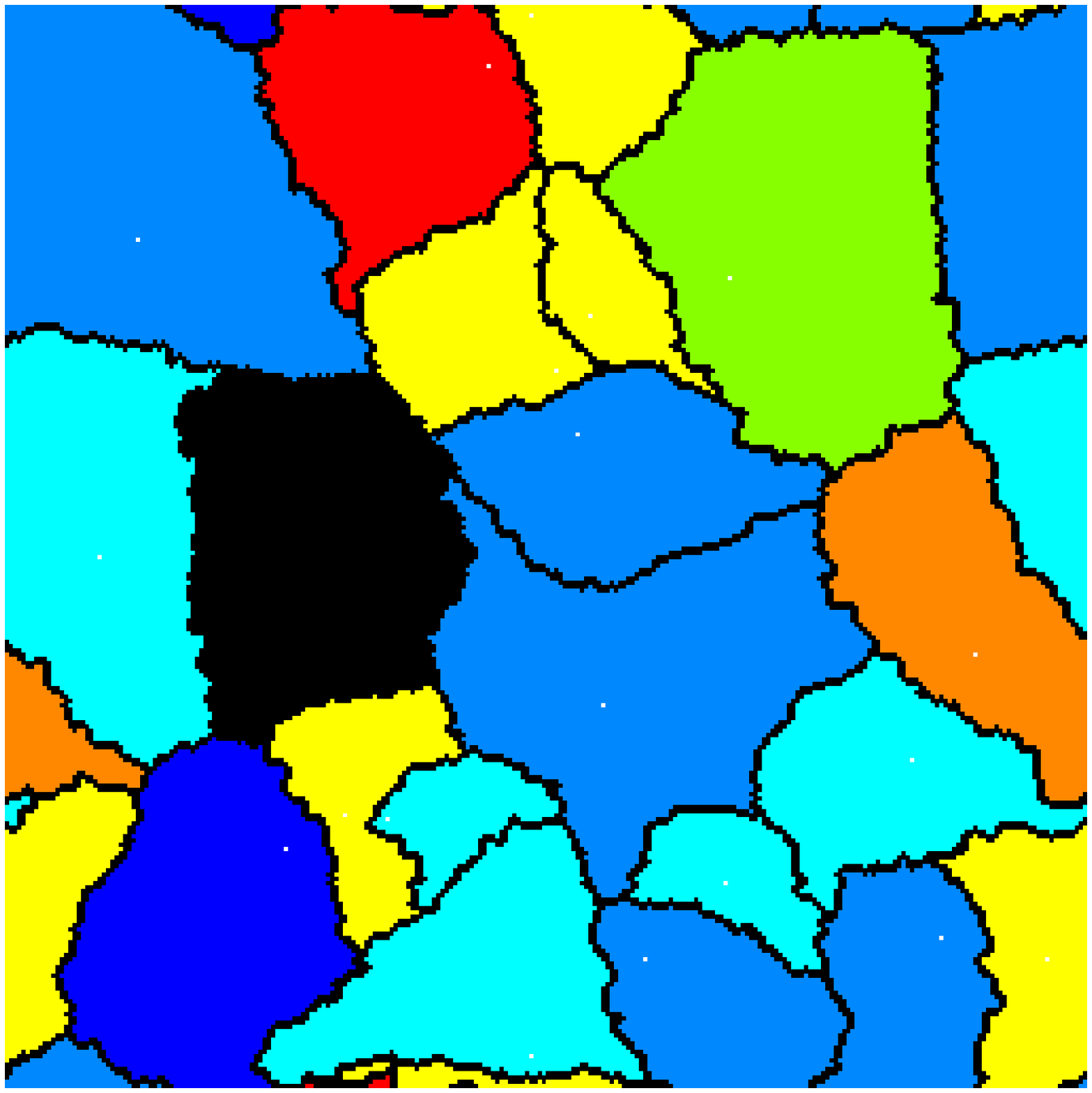}
\hfil
    \includegraphics[height=0.4\textwidth]{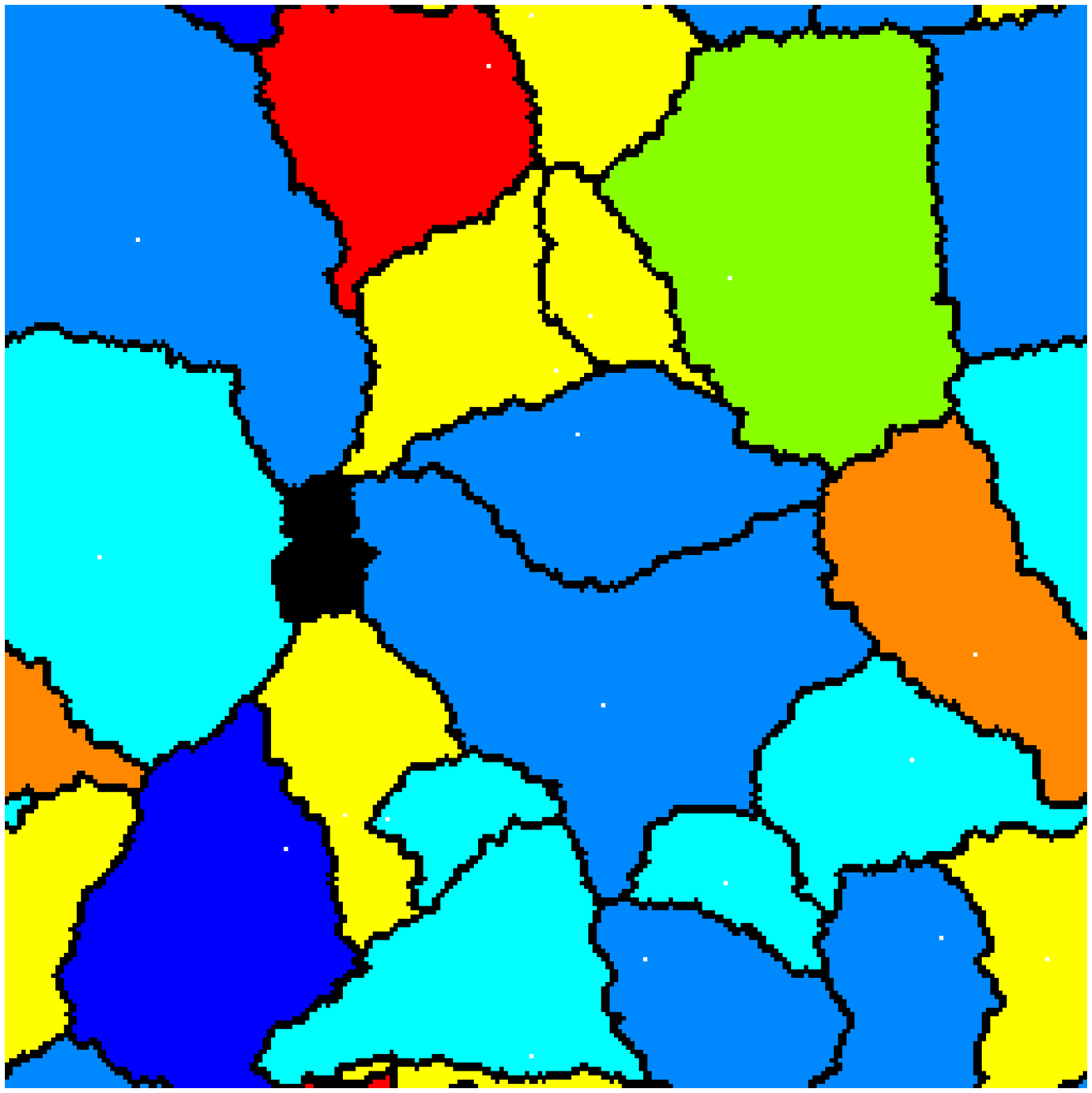}
}

    \caption{Snapshot of basic domain coarsening process.  LEFT: The
      black space comes from a cluster that has just been deleted.
      RIGHT: The black space is being taken over by the neighbors.
      --- Colors/grayscales are used to help the eye; clusters which
      have the same color/grayscale are still different clusters.
      System size $256^2$.}
    \label{fig:basic-snapshot}
  \end{center}
\end{figure}

In physics this is called a domain coarsening scheme
(e.g.~\cite{Flyvbjerg:foams}): Clusters are selected and deleted, and
their area is taken over by the neighbors.  This happens with a total
separation of time scales, that is, we do not pick another cluster for
deletion before the distribution of the last deleted cluster has
finished.  Fig.~\ref{fig:basic-snapshot} shows an example.  We will
call a cluster of size larger than zero ``active''.

Note that it is possible to pick a cluster that has already been
deleted.  In that case, nothing happens except that the clock advances
by one.  This implies that there are two reasonable definitions of
time:\begin{itemize}
  
\item \textbf{Natural time} $t$: This is the definition that we have
  used above.  In each time step, the probability of any given cluster
  to be picked for deletion is a constant $1/N$, where $N = L^2$ is
  the system size.  Note that it is possible to pick a cluster of size
  zero, which means that nothing happens except that time advances by
  one.
  
\item \textbf{Cluster time} $\tilde t$: An alternative is to chose
  between the \emph{active} clusters only.  Then, in each time step,
  the probability of any given cluster to be picked for deletion is
  $1/n(\tilde t)$, where $n(\tilde t) = N - \tilde t$ is the number of
  remaining active clusters in the system.

\end{itemize}
Although the dynamics can be described more naturally in cluster time,
we prefer natural time because it is closer to our economics
interpretation.

\begin{figure}[t]
  \begin{center}
\hbox{
    \includegraphics[width=0.49\hsize]{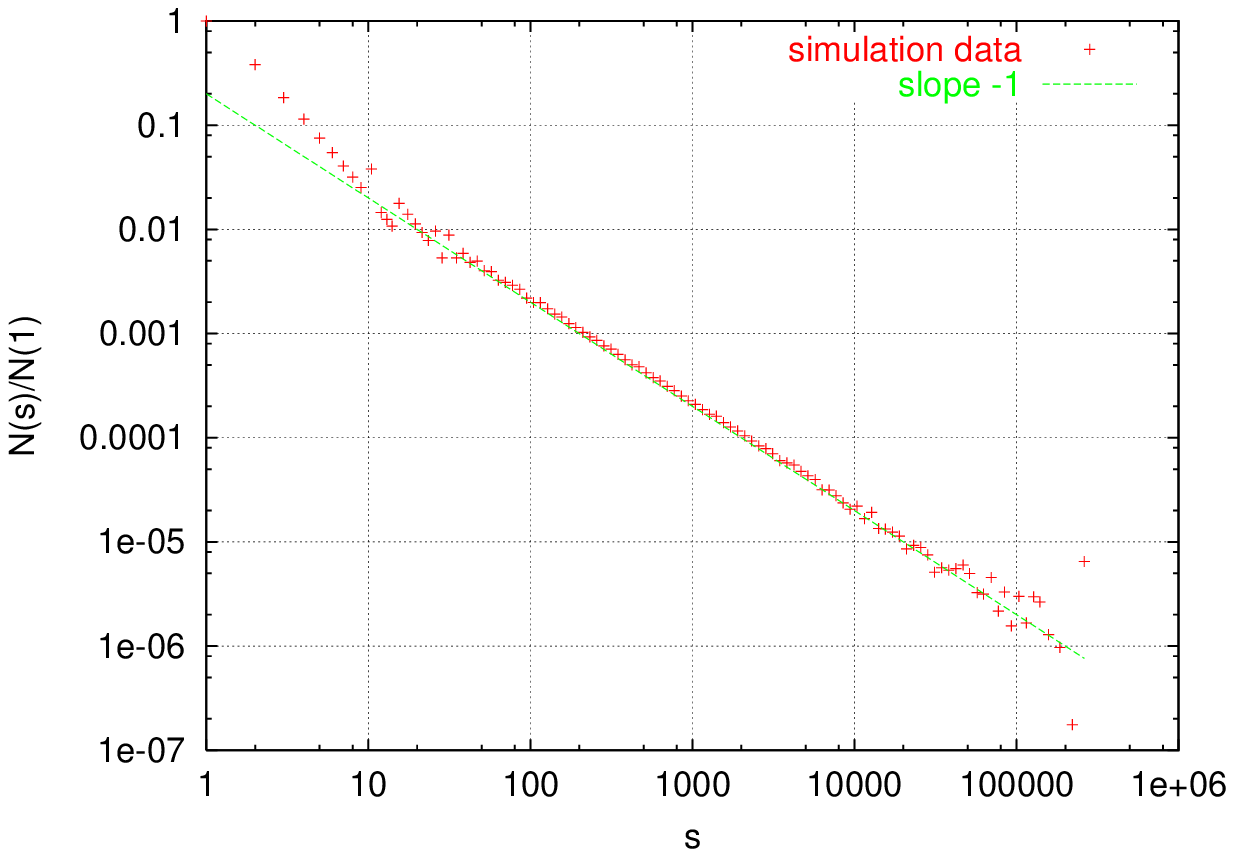}
\hfill
    \includegraphics[width=0.49\hsize]{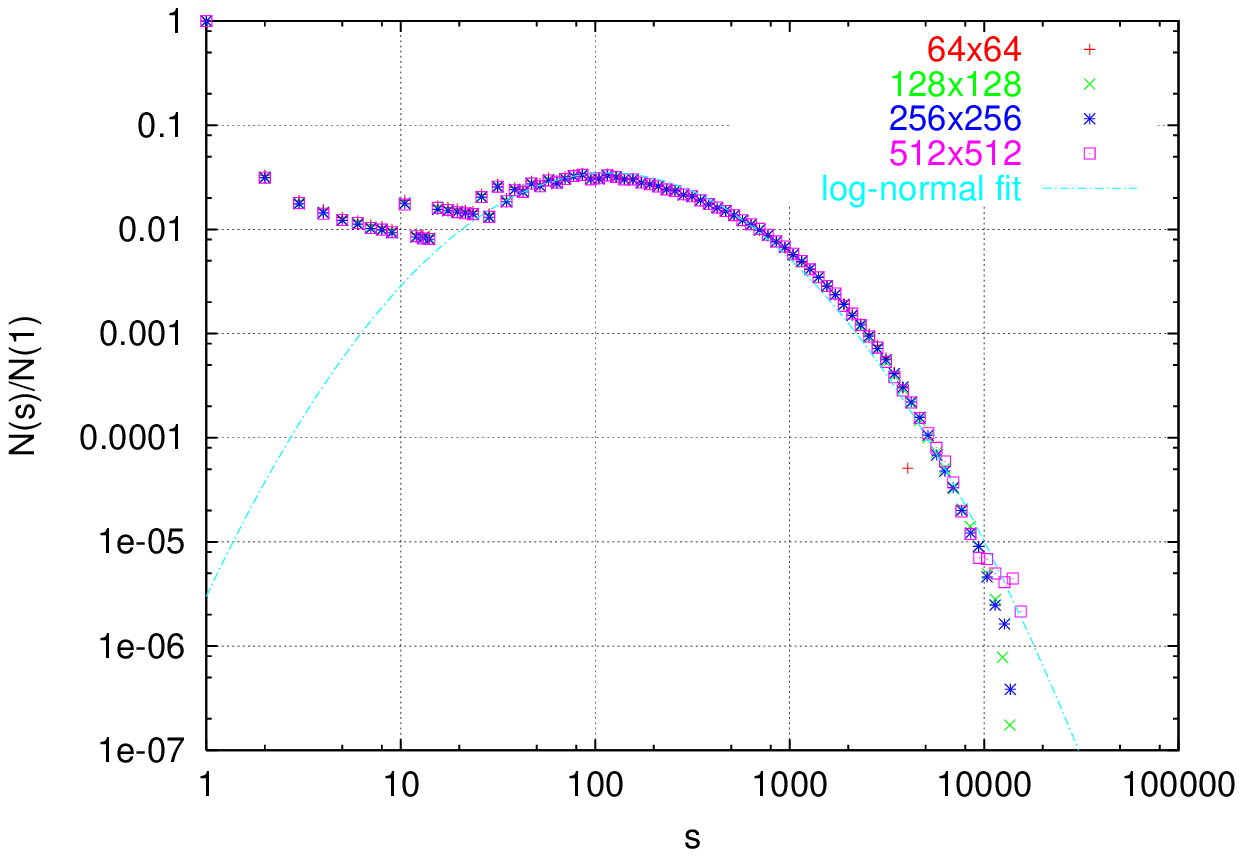}
}
    \caption{%
      LEFT: Cluster size distribution of the basic model without
      injection, in natural time.  Number of clusters per logarithmic
      bin, divided by number of clusters in first bin.  The straight
      line has slope $-1$, corresponding to $n(s) \sim s^{-2}$ because
      of logarithmic bins.  System size $512^2$.  As explained in the
      text, this is not a steady state distribution, but a
      distribution which emerges when averaging over the complete
      evolution from $N$ clusters of size one to one cluster of size
      $N$. RIGHT: Cluster size distribution for random injection.
      Number of clusters per logarithmic bin, divided by number of
      clusters in first bin.  The plot shows $p_{\it inj} = 0.01$ and
      system sizes $64^2$, $128^2$, $256^2$, and $512^2$.  The line is
      a log-normal fit.  This is a steady state distribution.
}
    \label{fig:basic-scaling}
  \end{center}
\end{figure}

At any particular time step, there is a typical cluster size.  In
fact, in cluster time, since there are
$n(\tilde t) = N - \tilde t$ clusters, the average cluster size as a
function of cluster time is
$\overline S(\tilde t) = N / n(\tilde t) = 1 / (1 - \tilde t/N)$.
However, if one \emph{averages over all time steps}, we find a scaling
law.  In cluster time, it is numerically close to
$
\label{domain:coarsening}
\tilde n(s) \sim s^{-3}   \hbox{ or } \tilde n(>\!s) \sim s^{-2} \ ,
$
where $s$ is the cluster size, $n(s)$ is the number of clusters of
size $s$, and $n(>\!s)$ is the number of clusters with size larger than
$s$.\footnote{%
  In this paper, we will also use $N(s) = s \, n(s)$ for the cluster
  size distribution in logarithmic bins, in particular for the
  figures. 
}  In natural time, the large clusters have more weight since time
moves more slowly near the end of the coarsening process.  The result
is again a scaling law (Fig.~\ref{fig:basic-scaling}~(left)), but with
exponents increased by one:
\[
\label{domain:coarsening:natural}
n(s) \sim s^{-2}   \hbox{ or } n(>\!s) \sim s^{-1} \ .
\]
It is important to note that this is not a steady state result.  The
result emerges when averaging over the whole time evolution, starting
with $N$ clusters of size one and ending with one cluster of size $N$.

%%%%%%%%%%%%%%%%%%%%%%%%%%%%%%%%%%%%%%%%%%%%

%%%%%%%%%%%%%%%%%%%%%%%%%%%%%%%%%%%%%%%%%%%%%
%%%%%%%%%%%%%%%%%%%%%%%%%%%%%%%%%%%%%%%%%%%%%
%\subsection{Foams}
%  
%I started looking into foams/bubbles, since they seem to be similar.
%But I couldn't find anything that was similar enough.  In particular,
%the usual mechanism in foams is that there is pressure on the small
%bubbles, and for that reason they lose air until they are gone.  In
%the model here, the proba to be eaten up is the same for all sizes. 

%%%%%%%%%%%%%%%%%%%%%%%%%%%%%%%%%%%%%%%%%%%%
%%%%%%%%%%%%%%%%%%%%%%%%%%%%%%%%%%%%%%%%%%%%
\subsection{Random injection with space}
\label{sec:rnd}

In view of evolution, for example in economics or in biology, it is
realistic to inject new small clusters.  A possibility is to inject
them at random positions.  So in each time step, before the cluster
deletion described above, in addition with probability $p_{\it inj}$
we pick one random site $i$ and inject a cluster of size one at $i$.
That is, we set $c(i)$ to $i$.  This is followed by the usual cluster
deletion.  It will be explained in more detail below what this means
in terms of system-wide injection and deletion rates.

This algorithm maintains the total separation of time scales between
the cluster deletion (slow time scale) and cluster growth (fast time
scale).  That is, no other cluster will be deleted as long as there
are still ``dead'' sites in the system.  Note that the definition of
time in this section corresponds to natural time.
  
The probability that the injected cluster is really new is reduced by
the probability to select a cluster that is already active.  The
probability of selecting an already active cluster is $n(t)/N$, where
$n(t)$ is again the number of active clusters. In consequence, the
effective injection rate is
\[
r_{\it inj,eff} = p_{\it inj} - n(t)/N \ .
\]
Similarly, the effective cluster deletion depends on the probability
of picking an active cluster, which is $n(t)/N$.  In consequence, the
effective deletion rate is
\[
r_{\it del,eff} = n(t)/N \ .
\]
This means that, in the steady state, there is a balance of injection
and deletion, $n_*/N = p_{\it inj} - n_* / N$, and thus the steady state
average cluster number is
\[
n_* = N \, p_{\it inj} / 2 \ .
\]
In consequence, the steady state average cluster size is
\[
s_* = N/n_* = 2 / p_{\it inj} \ .
\]

The cluster size distribution for the model of this section is
numerically close to a log-normal distribution, see
Fig.~\ref{fig:basic-scaling}~(right).  Indeed, the position of the
distribution moves with $1/p_{\it inj}$ (not shown).  In contrast to
Sec.~\ref{sec:basic}, this is now a steady state result. 

%%%%%%%%%%%%%%%%%%%%%%%%%%%%%%%%%%%%%%%%%%%%
%%%%%%%%%%%%%%%%%%%%%%%%%%%%%%%%%%%%%%%%%%%%
\subsection{Injection on a line}

\begin{figure}[t]
  \begin{center}
\centerline{
    \includegraphics[width=0.4\textwidth]{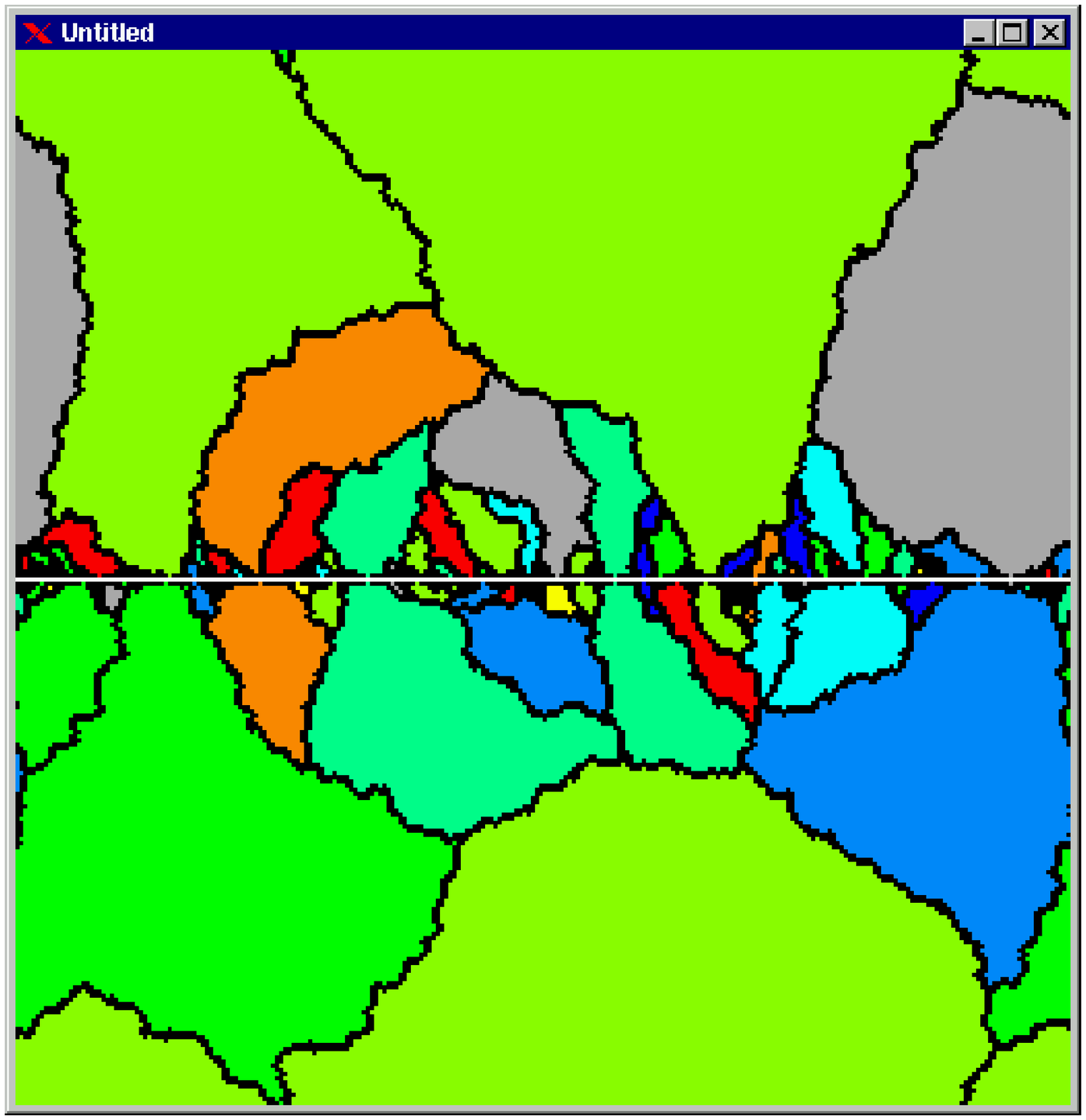}
\hfil
    \includegraphics[width=0.55\textwidth]{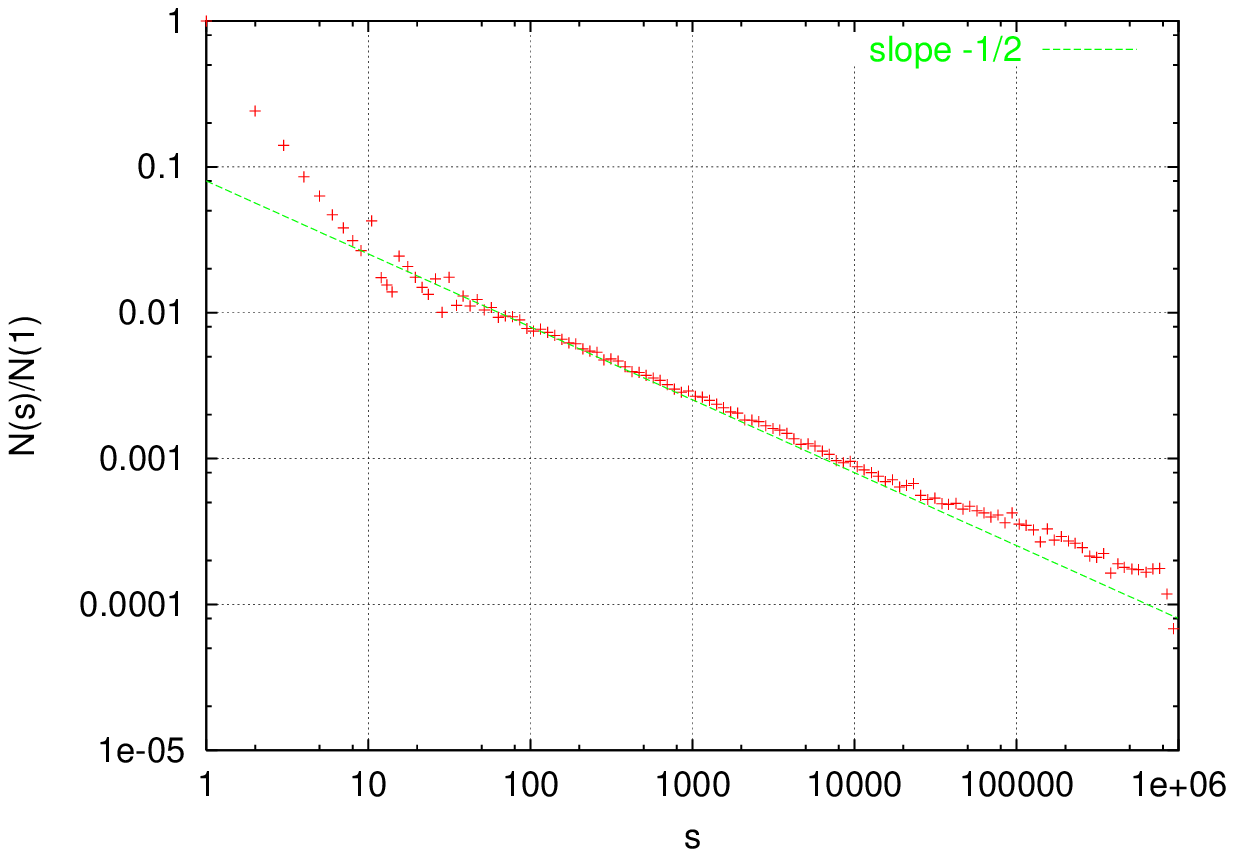}
}
    \caption{%
      LEFT: Injection along a line.  System size $256^2$.  RIGHT:
      Scaling plot for basic model plus injection on a line.  Number
      of clusters per logarithmic bin, divided by number of clusters
      in first bin.  The straight line has slope $-1/2$ corresponding
      to $n(s) \sim s^{-3/2}$.  System size $1024^2$.  This is a
      steady state distribution.
}
    \label{fig:snap:line}
  \end{center}
\end{figure}

It is maybe intuitively clear that the injection mechanism of the
model described in Sec.~\ref{sec:rnd} destroys the scaling law from
the basic model without injection (Sec.~\ref{sec:basic}), since
injection at random positions introduces a typical spatial scale.  One
injection process that actually generates steady-state scaling is
injection along a 1-d line.  Instead of the random injection of
Sec.~\ref{sec:rnd}, we now permanently set
\[
  c(i) = i
\]
for all sites along a line.  Fig.~\ref{fig:snap:line}~(left) shows a snapshot
of this situation.

%\yyyy{what about Maya's remark that it there is a difference if the
%  root of the deleted cluster can grow back in the same time step or
%  not?  I think I tried it and it didn't matter.}

In this case, we numerically find a stationary cluster size
distribution (Fig.~\ref{fig:snap:line}~(right)) with
\[
n(s) \sim s^{-1.5}   \hbox{ or }  n(>\!s) \sim s^{-0.5} \ .
\]
Since the injection mechanism here does not depend on time, and since
the cluster size distribution itself is stationary, it is independent
from the specific definition of time.

%%%%%%%%%%%%%%%%%%%%%%%%%%%%%%%%%%%%%%%%%%%%

%%%%%%%%%%%%%%%%%%%%%%%%%%%%%%%%%%%%%%%%%%%%
%%%%%%%%%%%%%%%%%%%%%%%%%%%%%%%%%%%%%%%%%%%%
\subsection{Random injection without space}
\label{sec:non-sptl}

\begin{figure}[t]
  \begin{center}
\centerline{%
    \includegraphics[width=0.49\hsize]{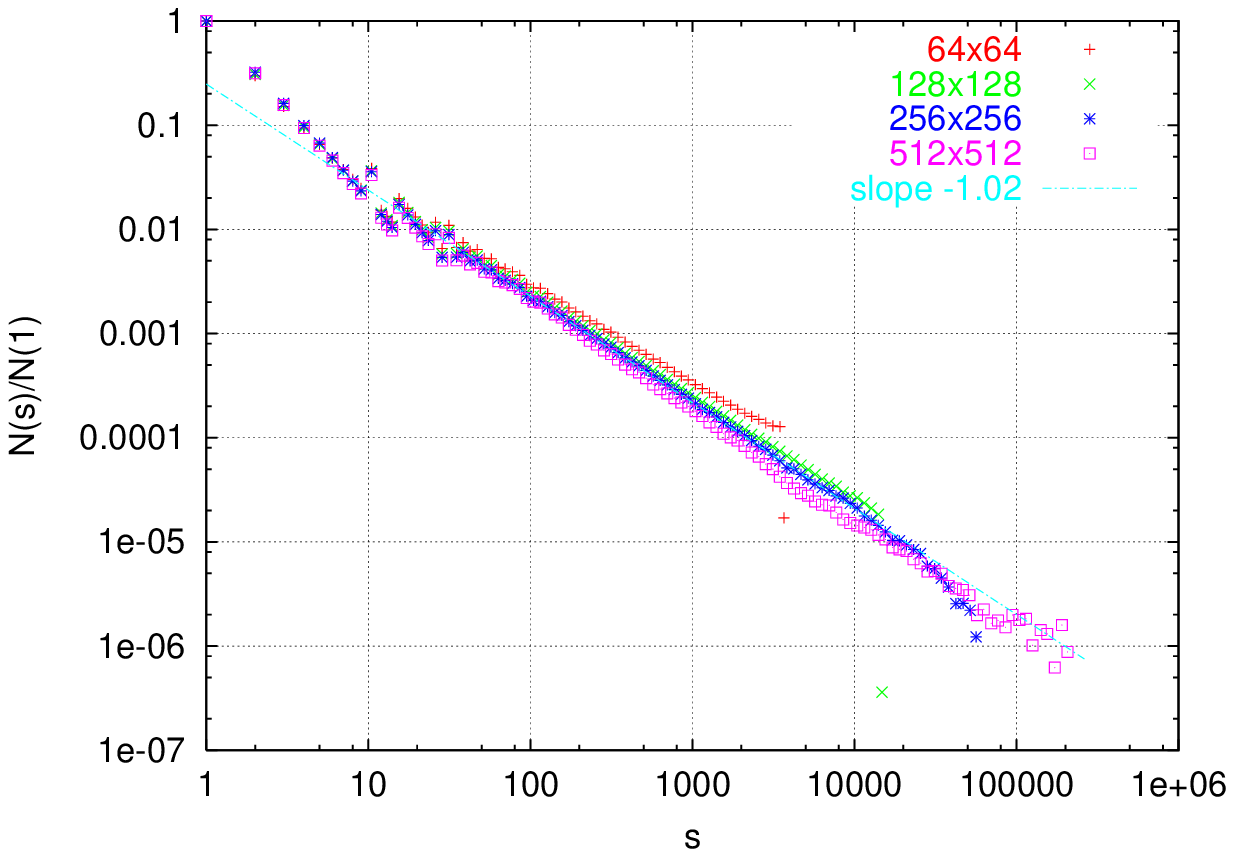}
\hfil
    \includegraphics[width=0.49\hsize]{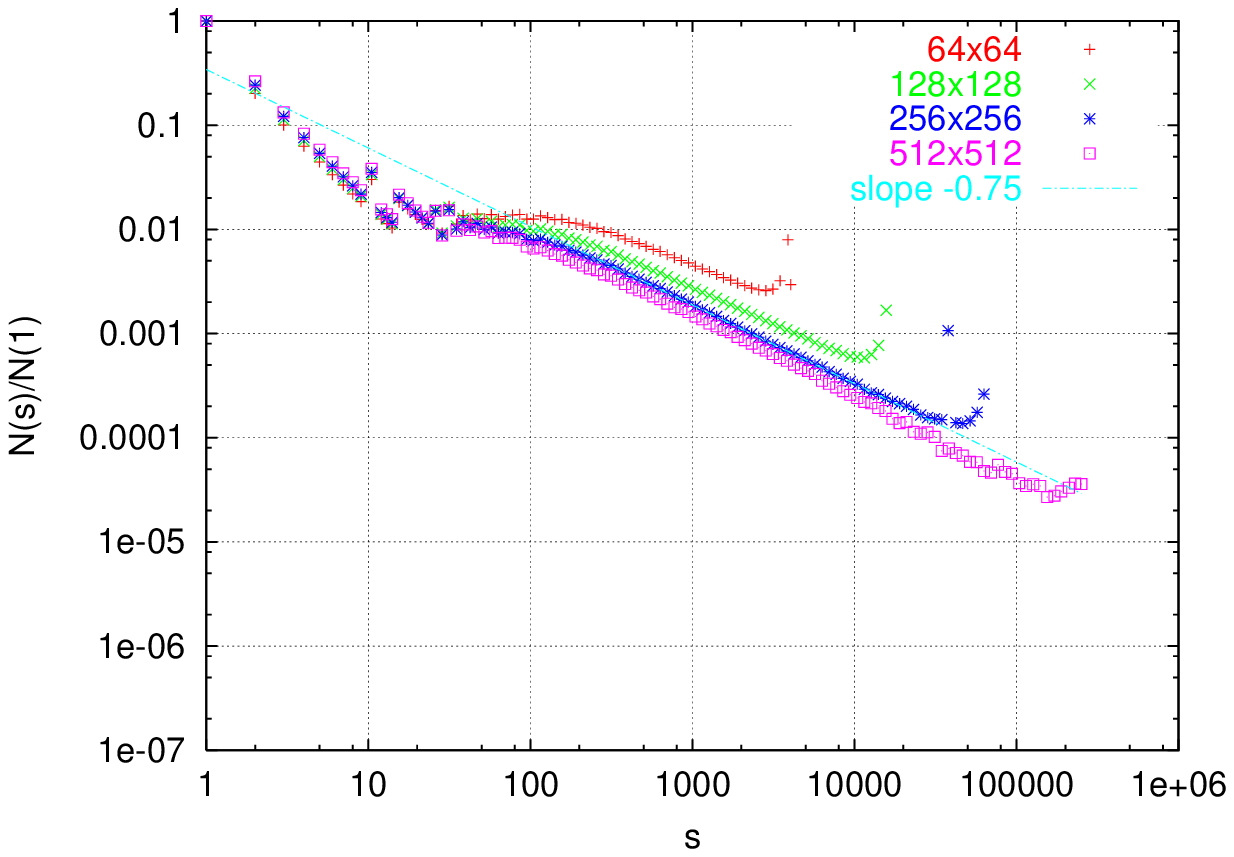}
}
    \caption{%
      Steady state cluster size distributions for different
      non-spatial simulations.  Number of clusters per logarithmic
      bin, divided by number of clusters is first bin.  System sizes
      $64^2$ to $512^2$.  LEFT: $p_{\it inj}=0.1$.  RIGHT: $p_{\it
        inj}=0.01$.
}
    \label{fig:non-spatial-gpl}
  \end{center}
\end{figure}

One could ask what would happen without space.  A possible translation
of our model into ``no space'' is: Do in parallel: Instead of picking
one of your four nearest neighbors, you pick an arbitrary other agent
(random neighbor approximation).  If that agent is not dead, copy its
cluster number. Do this over and over again in parallel, until all
agents are part of a cluster again.  A cluster is now no longer a
spatially connected structure, but just a set of agents.  In that
case, we obtain again power laws for the size distribution, but this
time with slopes that depend on the injection rate~$p_{\it inj}$
(Fig.~\ref{fig:non-spatial-gpl}); see Sec.~\ref{sec:theo-wo-space} for
details.

%%%%%%%%%%%%%%%%%%%%%%%%%%%%%%%%%%%%%%%%%%%%

%%%%%%%%%%%%%%%%%%%%%%%%%%%%%%%%%%%%%%%%%%%%
%%%%%%%%%%%%%%%%%%%%%%%%%%%%%%%%%%%%%%%%%%%%
%%  \subsection{Other spatial dimensions}
%%  
%%  One can make versions of the above models for other dimensions.  1-d
%%  is somewhat degenerated since we need to decide if injected small
%%  clusters cut the large cluster into two pieces or not.  In 2-d or
%%  higher, there is no problem here.

\subsection{Real world company size distributions}

\begin{figure}[t]
  \begin{center}
%    \ncludegraphics[height=0.25\textheight]{sumsales-loglog-gpl.eps}
%    \ncludegraphics[height=0.4\textheight]{sales-loglog-gpl.eps}
%    \ncludegraphics[height=0.4\textheight]{sales-ylog-gpl.eps}
%    \ncludegraphics[width=\hsize]{sales-xlog-gpl.eps}
%    \ncludegraphics[height=0.4\textheight]{sales-accum-gpl.eps}
\centerline{
    \includegraphics[width=0.49\textwidth]{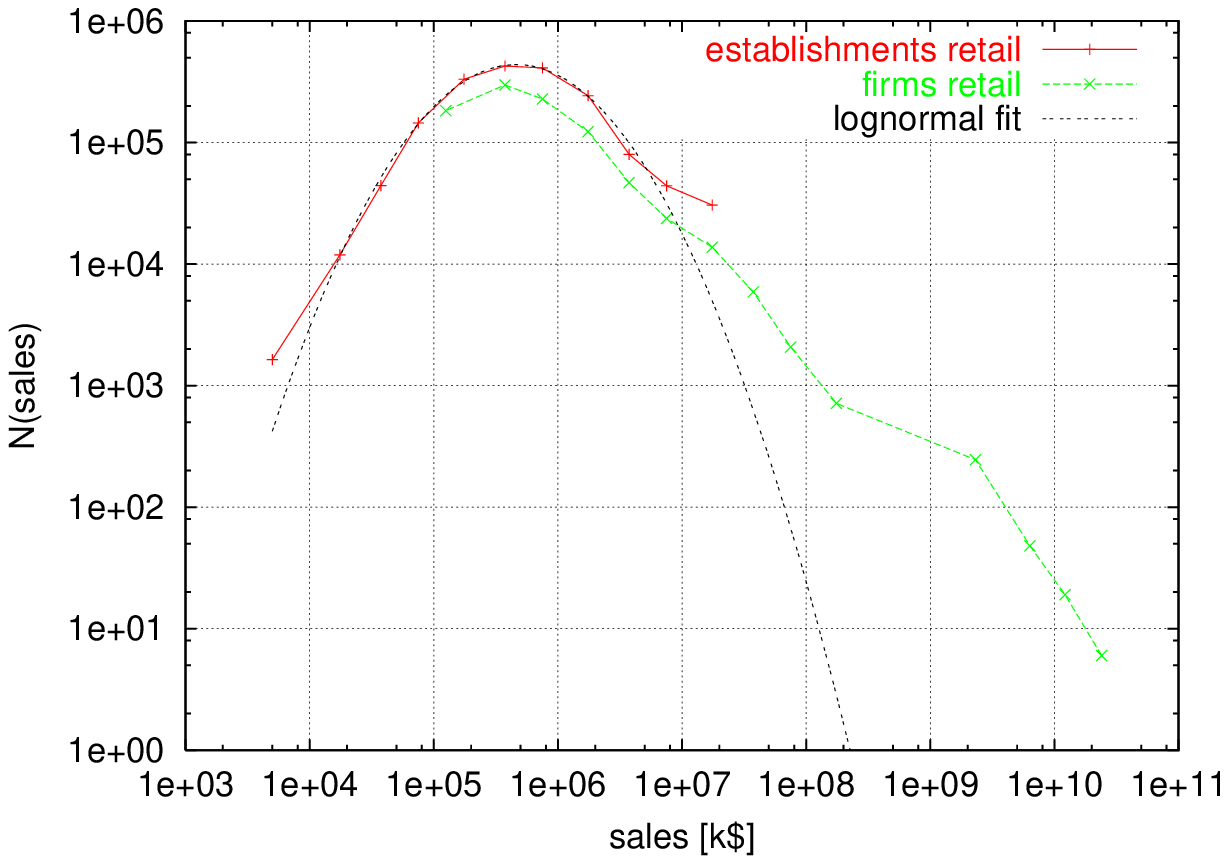}
\hfill
    \includegraphics[width=0.49\textwidth]{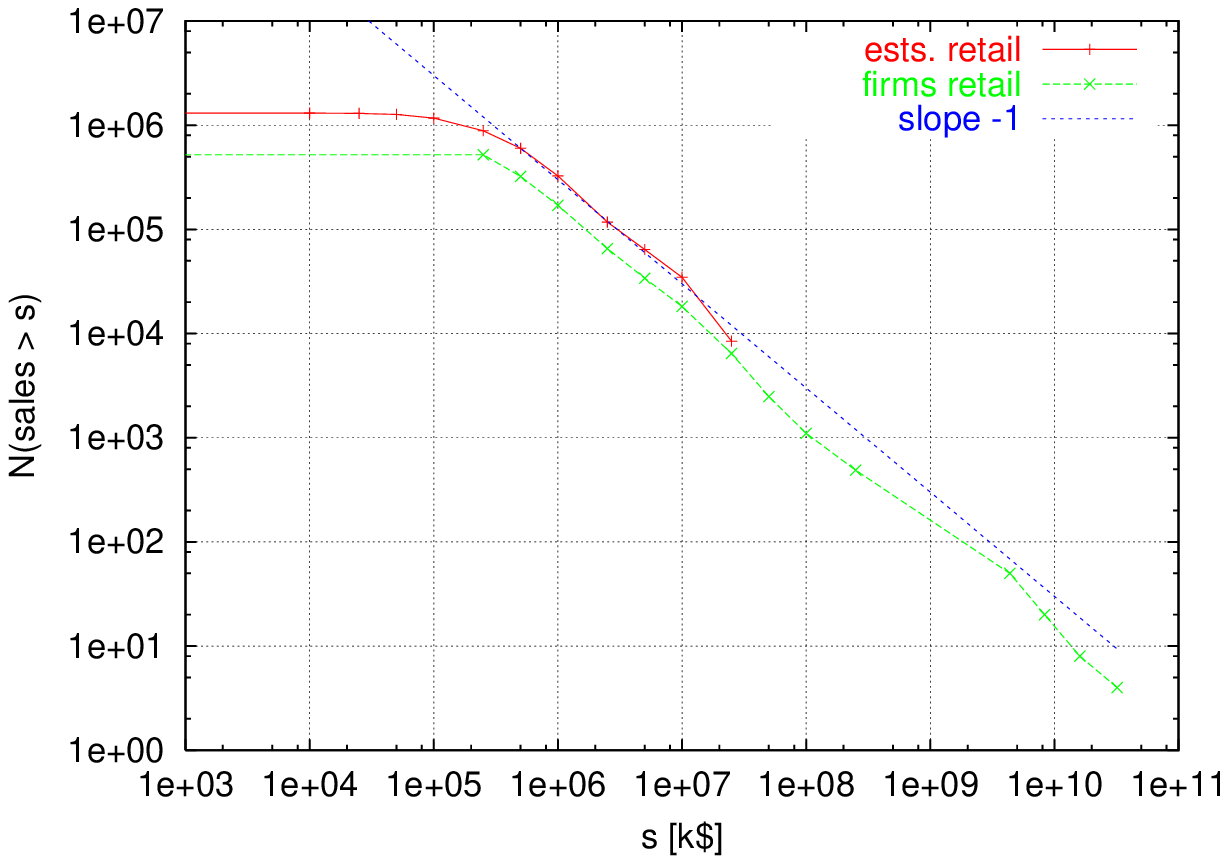}
}
    \caption{%
%
%TOP: Number of establishments exceeding certain sales. Log-log 
%      plot.  
%
      1992 U.S.\ Economic Census data.  LEFT: Number of retail
      establishments/retail firms per logarithmic bin as function of
      annual sales.  RIGHT: Number of establishments/firms which have
      more sales than a certain number.
}
    \label{fig:sales}
  \end{center}
\end{figure}

Fig.~\ref{fig:sales} shows actual retail company size distributions
from the 1992 U.S.\ economic census~\cite{econ:census:92}, using
annual sales as a proxy for company size.  We use the retail sector
because we think that it is closest to our modelling assumptions ---
this is discussed at the end of Sec.~\ref{sec:discussion}.  We show
two curves: establishment size,
and firm size.\footnote{%
An establishment is ``a single physical location at
which business is conducted.  It is not necessarily identical with a
company or enterprise, which may consist of one establishment or
more.''~\cite{econ:census:92}.
} It is clear that in order to be comparable with our model
assumptions, we need to look at establishment size rather than at
company size.

Census data comes in unequally spaced bins; the procedure to convert
it into useable data is described in the appendix.  Also, the last
four data points for firm size (not for the establishment size,
however) were obtained via a different method than the other data
points; for details, again see the appendix.

From both plots, one can see that there is a typical establishment
size around \$400\,000 annual sales; and the typical firm size is a
similar number.  This number intuitively makes sense: With, say,
income of 10\% of sales, smaller establishments will not provide a
reasonable income.

One can also see from the plots that the region around that typical
size can be fitted by a log-normal.  We also see, however, that for
larger numbers of annual sales, such a fit is impossible since the
tail is much fatter.  A scaling law with
\[
n(>\!s) \sim s^{-1} \hbox{\ \ corresponding to \ \ } n(s) \sim s^{-2}
\]
is an alternative here.\footnote{%
  Remember again, that slopes from log-log plots in logarithmic bins
  are different by one from the exponent in the distribution.  So
  $n(s) \sim s^{-2}$ corresponds to a slope $-1$ \emph{both} in the
  accumulated distribution $n(>\!s)$ and when plotting logarithmic bins
  $N(s)/N(1)$. 
}

This is, however, at odds with investigations in the literature.  For
example, Ref.~\cite{Stanley:sizes:econ-letters} find a log-normal, and
by using a Zipf plot they show that for large companies the tail is
\emph{less} fat than a log-normal.  However, there is a huge
difference between our and their data: They only use \emph{publicely
  traded} companies, while our data refers to all companies in the
census.  Indeed, one finds that their plot has its maximum at annual
sales of $\$10^8$, which is already in the tail of our distribution.
This implies that the small scale part of their distribution comes
from the fact that small companies are typically not publicely traded.
In consequence, it reflects the dynamics of companies entering and
exiting from the stock market, not entry and exit of the company
itself.

We conclude that from available data, company size distributions are
between a log-normal and a power law with $n(s) \sim s^{-2}$ or $n(>\!s)
\sim s^{-1}$.  Further investigation of this goes beyond the scope of
this paper.

%%  \begin{figure}[p]
%%    \begin{center}
%%  %    \ncludegraphics[height=0.25\textheight]{sumempl-loglog-gpl.eps}
%%      \includegraphics[height=0.4\textheight]{empl-loglog-gpl.eps}
%%  %    \ncludegraphics[height=0.4\textheight]{empl-ylog-gpl.eps}
%%  %    \ncludegraphics[width=\hsize]{empl-xlog-gpl.eps}
%%      \caption{%
%%  %
%%  %      TOP: Number of establishments exceeding certain number of
%%  %      employees.  Log-log plot.  
%%        Normalized number of establishments as a function of the number
%%        of employees. Normalized means that the number is divided by the
%%        class size.  Log-log plot.  
%%  %BOTTOM:
%%  %      Normalized number of establishments with certain number of
%%  %      employees. Semi-log plot.
%%  %
%%  }
%%      \label{fig:empl}
%%    \end{center}
%%  \end{figure}

%\processdelayedfloats
%%%%%%%%%%%%%%%%%%%%%%%%%%%%%%%%%%%%%%%%%%%%%%%%%%%%
\section{Theoretical considerations}
\label{sec:theo}

\subsection{Spatial coarsening model (slope -2 in natural time)}

We are looking again at the ``basic model''.  In cluster time this
was: randomly pick one of the clusters, and give it to the neighbors.
The following heuristic model gives insight:
\begin{enumerate}

\item We start with $N$ clusters of size 1.
  
\item We need $N/2$ time steps to delete $N/2$ of them and with that
  generate $N/2$ clusters of size~2.

\item In general, we need $N/2^k$ time steps to move from
  $N/2^{k-1}$ clusters of size $2^{k-1}$ to $N/2^k$ clusters of
  size $2^k$. 
  
\item If we sum this over time, then in each logarithmic bin at
  $s=2^k$ the number of contributions is $N/2^k \times N/2^k$, i.e.\ 
  $\sim s^{-2}$.

\item Since these are logarithmic bins, this corresponds to 
  $\tilde n(s) \sim s^{-3} \hbox{ \ \ or \ \ } \tilde n(>\!s) \sim s^{-2} \ , $ which
  was indeed the simulation result in cluster time.

\item
In natural time, we need a constant amount of time to move from $k-1$
to $k$, and thus obtain via the same argument
$
n(s) \sim s^{-2} \hbox{ \ \ or \ \ } n(>\!s) \sim s^{-1} \ ,
$
which was the simulation result in natural time.

\end{enumerate}

\subsection{Random injection in space (log-normal)}

At the moment, we do not have a consistent explanation for the
log-normal distribution in the spatial model.  A candidate is the
following: Initially, most injected clusters of size one are
\emph{within} the area of some larger and older cluster.  Eventually,
that surrounding cluster gets deleted, and all the clusters of size
one spread in order to occupy the now empty space.  During this phase
of fast growth, the speed of growth is proportional to the perimeter,
and thus to $\sqrt{s}$, where $s$ is the area.  Therefore, $\sqrt{s}$
follows a biased multiplicative random walk, which means that
$\log(\sqrt{s}) = \log(s)/2$ follows a biased additive random walk.
In consequence, once that fast growth process stops, $\log(s)$ should
be normally distributed, resulting in a log-normal distribution for
$s$ itself.  In order for this to work, one needs that this growth
stops at approximately the same time for all involved clusters.  This
is apprixomately true because of the ``typical'' distance between
injection sites which is inversely proportional to the injection rate.
More work will be necessary to test or reject this hypothesis.

\subsection{Injection on a line (slope -3/2)}

If one looks at a snapshot of the 2D picture for ``injection on a
line'' (Fig.~\ref{fig:snap:line}), one recognizes that one can
describe this as a structure of cracks which are all anchored at the
injection line.  There are $L$ such cracks (some of length zero); 
cracks merge with increasing distance from the injection line, but
they  do not branch.

According to Ref.~\cite{Sneppen:etc:fractal-cracks}, this leads naturally
to a size exponent of $-3/2$, as found in the simulations.  The
argument is the following: The whole area, $L^2$, is covered by
\[
\int ds \, s \, n(s) \ ,
\]
where $n(s)$ is the number of clusters of size $s$ on a linear scale.
We assume $n(s) \sim s^{-\tau}$, however the normalization is missing.
If all clusters are anchored at a line of size $L$, then a doubling of 
the length of the line will result in twice as many clusters.  In
consequence, the normalization constant is $\propto L$, and thus 
$n(s) \sim L \, s^{-\tau}$.  Now we balance the total area, $L^2$,
with what we just learned about the covering clusters:
\[
L^2 \sim \int ds \, s \, L \, s^{-\tau}
= L \, \int ds \, s^{1-\tau}
\sim L \, s^{2-\tau} \big|_0^S \ .
\]
%% Because of the lower integration boundary, this works only for 
%% $\tau < 2$.

Assuming that $\tau < 2$, then the integral does not converge for 
$S \to \infty$, and we need to take into account how the cut-off $S$
scales with $L$.  This depends on how the cracks move in space as a
function of the distance from the injection line. If the cracks are
roughly straight, then the size of the largest cluster is $\sim L^2$.
If the cracks are random walks, then the size of the largest cluster
is $\sim L^{3/2}$.  In consequence:\begin{itemize}

\item For ``straight'' lines:  $L^2 \sim L \; (L^2)^{2-\tau}$
  $\ \Rightarrow\ $ 
$
2 = 1 + 2 \, (2-\tau) %= 1 + 4 - 2 \tau
$
$\ \Rightarrow\ $
$
\tau = 3/2 \ .
$

\item For random walk:
$
2 = 1 + 3/2 \, (2-\tau) % = 1 + 3 - 3 \tau/2
 \ \ \Rightarrow \ \ 
\tau = 4/3 \ .
$

\end{itemize}
Since our simulations result in $\tau \approx 3/2$, we conclude that
our lines between clusters are not random walks.  This is intuitively
reasonable: When a cluster is killed, then the growth is biased
towards the center of the deleted cluster, thus resulting in random
walks which are all differently biased.  This bias then leads to the
``straight line'' behavior.
--- This implies that the $\sim s^{-3/2}$ steady state scaling law 
hinges on two ingredients (in a 2D system):
(i) The injection comes from a 1D structure.  
(ii) The boundaries between clusters follow something that
  corresponds to straight lines.  As we have seen, the biasing of a
  random walk is already enough to obtain this effect.

%%  With respect to economics, in our view it remains an open question if
%%  this could be an explanation for the fat tail discrepancies from the
%%  log-normal of the establishment size distributions.  We believe that
%%  the ``story'' itself is somewhat plausible: Far away from the ``big
%%  city'', customers are bound to not know more than one or two stores
%%  there, and more likely than not, they will have that information from
%%  their neighbors.  It is also plausible that the number of stores grows
%%  more slowly than the number of customers $N$: This would simply mean
%%  that when population density grows, we have a tendency to get bigger
%%  stores instead of more of them.\footnote{%
%%  %
%%    Quite possibly, some version of spatial impedance (transportation
%%    cost) could generate such a dependency from more microscopic
%%    principles.
%%  %
%%  }

\subsection{Injection without space (variable slope)}
\label{sec:theo-wo-space}

Without space, clusters do not grow via neighbors, but via random
selection of one of their members.  That is, we pick a cluster, remove
it from the system, and then give its members to the other clusters
one by one.  The probability that the agent choses a cluster $i$ is
proportional to that cluster's size $s_i$.  If for the moment we
assume that time advances with each member which is given back, we
obtain the rate equation
\[
{d n(s) \over dt}
= (s\!-\!1) \,  n(s\!\!-\!\!1) - \epsilon \, n(s) - s \,  n(s)
- \epsilon \, p_{\it inj} \, n(s)
+ \epsilon \, p_{\it inj} \, n(s\!\!+\!\!1) \ .
\]
The first and second term on the RHS represent cluster growth by
addition of another member; the third term represents random deletion;
the fourth and fifth term the decrease by one which happens if one of
the members is converted to a start-up via injection.  $\epsilon$ is
the rate of cluster deletion; since we first give all members of a
deleted cluster back to the population before we delete the next
cluster, it is proportional to the inverse of the average cluster size
and thus to the injection rate:
$\epsilon \sim 1/\langle s \rangle \sim p_{\it inj}$.  This is similar to
an urn process with additional deletion.

Via the typical approximations
$s \, n(s) - (s-1) \, n(s-1) \approx {d \over ds} (s \, n(s))$ etc.\ we
obtain, for the steady state, 
the differential equation
\[
0 = - N - s \, {dN \over ds} - \epsilon \, N 
+ \epsilon \, p_{\it inj} \, {dN \over ds} \ .
%
%dN/N = - (1+\epsilon) \, ds / (s - \epsilon p_{\it inj}) \ .
\]
This leads to
\[
n(s) \propto (s - \epsilon p_{\it inj})^{-(1+\epsilon)}
\sim s^{-(1+\epsilon)} \ .
\]
That is, the exponent depends on the injection rate, and in the limit
of $p_{\it inj} \to 0$ it goes to $-1$.  This is indeed the result
from Sec.~\ref{sec:non-sptl} (see
Fig.~\ref{fig:non-spatial-gpl}).\footnote{%
  Note that the approach in this section corresponds to measuring the
  cluster size distribution every time we give an agent back to the
  system, while in the simulations we measured the cluster size
  distribution only just before a cluster was picked for deletion.  In
  how far this is important is an open question; preliminary
  simulation results indicate that it is important for the spatial
  case with injection but not important for the non-spatial case in
  this section.
}

%%%%%%%%%%%%%%%%%%%%%%%%%%%%%%%%%%%%%%%%%%%%
\section{Price formation}
\label{sec:prices}

What we will do now is to add the mechanism of price formation to our
spatial competition model.  For this, we identify sites with
consumers/customers.  Clusters correspond to domains of consumers who
go to the same shop/company.  Intuitively, it is clear how this should
work: Companies which are not competitive will go out of business, and
their customers will be taken over by the remaining companies.  The
reduction in the number of companies is balanced by the injection of
start-ups.
Companies can go out of business for two reasons: losing too much
money, or losing too many customers.  The first corresponds to a price
which is too low; the second corresponds to a price which is too high.

We model these aspects as follows: We again have $N$ sites on an $N= L
\times L$ grid with periodic boundary conditions (torus).  On each
site, we have a consumer and a firm.  These are not connected in any
way except by the spatial position -- one can imagine that the firm is
located ``downstairs'' while the consumer lives ``upstairs''.  Firms
with customers are called ``active'', the other ones ``inactive''. A
time step consists of the following
sub-steps:\begin{itemize}

\item Trades are executed.

\item Companies with negative profit go out of business.
  
\item Companies change prices.

\item New companies are injected.

\item Consumers can change where they shop.

\end{itemize}
These steps are described in more detail in the following:

\textbf{Trade:} All customers have an initial amount $M$ of money,
which is completely spent in each time step and replenished in the
next.  Every customer $i$ also knows which firm $j=f(i)$ he/she buys
from.  Thus, he/she orders an amount $Q_i = M/P_j$ at his/her company,
where $P_j$ is that company's price.  The companies produce to order,
and then trades are executed.  That is, a company that has $n_j$
customers and price $P_j$ will produce and sell
$Q_j = n_j M/P_j$ units and will collect $n_j M$ units of money.

\textbf{Company exit:} We assume an externally given cost function for
production, $C(Q)$, which is the same for everybody.  If profit $\Pi_j
:= n_j \, M - C(n_j \, M/P_j)$ is less than zero, then the company is
losing
money and will immediately go out of business.\footnote{%
In this model no accumulation of assets is allowed.  This
simplification will be relaxed in future work.
} The prices of such a company is set to infinity.  We will use
$C(Q)=Q$, corresponding to a linear cost of production.  With this
choice, companies with prices $P_j<1$ will exit according to this rule
as soon as they attract at least one customer.

\textbf{Price changes:} With probability one, pick a random integer
number between $1$ and $N$.  If there is an active company with that
number, its price is randomly increased or decreased by $\delta$.

\textbf{Company injection:} Companies are made active by giving them
one customer: With probability $p_{\it inj}$, pick a random site $i$ and
make the consumer $i$ go shopping at company $i$.  The price of the
injected company is set to the price that the customer has paid
before, randomly increased or decreased by $\delta$.

% If I don't use the $\delta$ here, then prices cannot go up. 

\textbf{Consumer adaptation:} 
All customers whose prices got increased (either via ``company exit''
or via ``price changes'') will search for a new shop.\footnote{%
  The simplification that customers react to price changes only is
  useful because it leads to the separation of time scales between
  consumer behavior and firm behavior.
  } These ``searching'' consumers correspond to dead sites in the
basic spatial models (Sec.~3), and the dynamics is essentially a
translation of that: All searching consumers in parallel pick a random
nearest neighbor.  If that neighbor is also searching, nothing
happens.  If that neighbor is however not searching, and if that
neighbor is paying a lower price, our consumer will accept the
neighbor's shop.  Otherwise the customer will remain with her old
shop, and she will no longer search.  We keep repeating this until no
consumer is searching any more.

This model does not invest much in terms of rational or organized
behavior by any of the entities.  Firms change prices randomly; and
they exit without warning when they lose money.  New companies are
injected as small variations of existing companies.  Consumers only
make moves when they cannot avoid it (i.e.\ their company went out of
business and they need a new place to go shopping) or when prices just
went up.  Only in the last case they actively compare some prices.  It
will turn out (see below) that even that price comparison is not
necessary.

\begin{figure}[t]
  \begin{center}
\centerline{
   \includegraphics[width=0.49\textwidth]{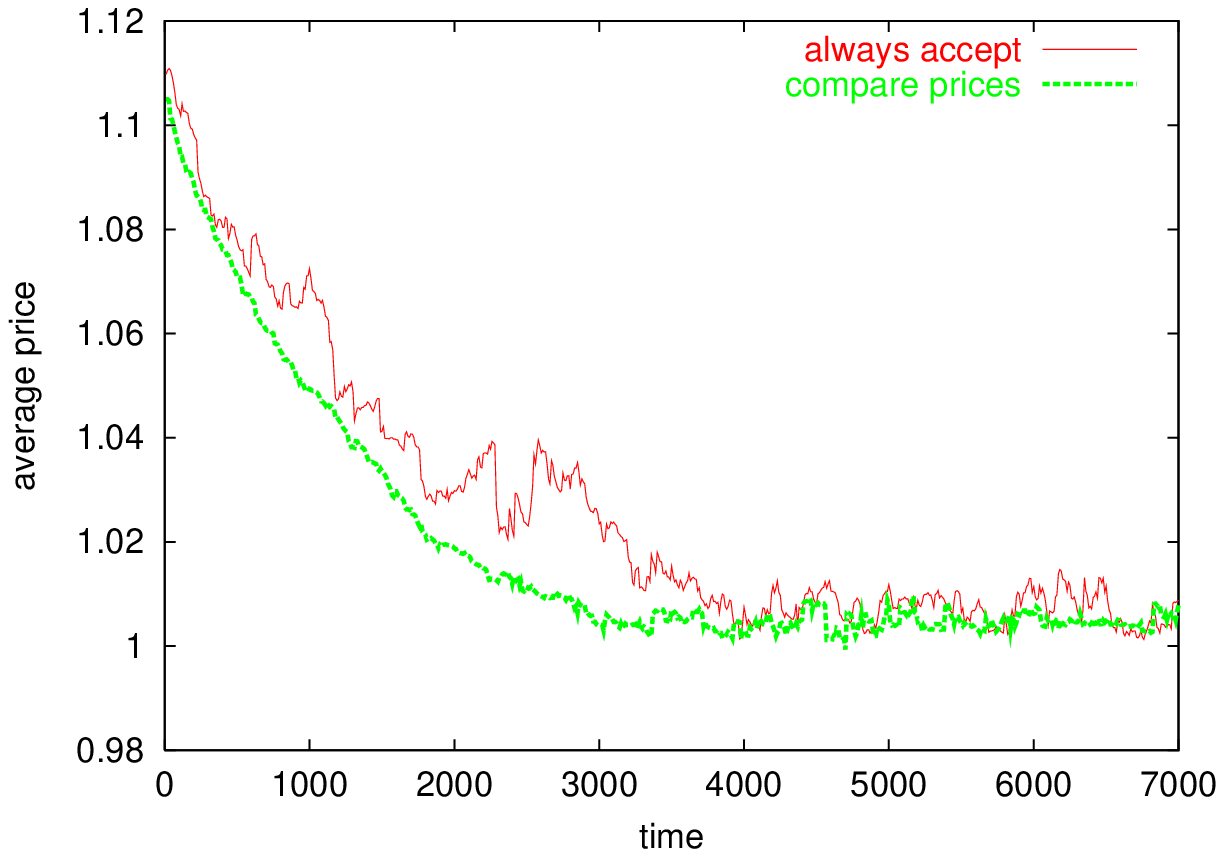}
\hfill
    \includegraphics[width=0.49\textwidth]{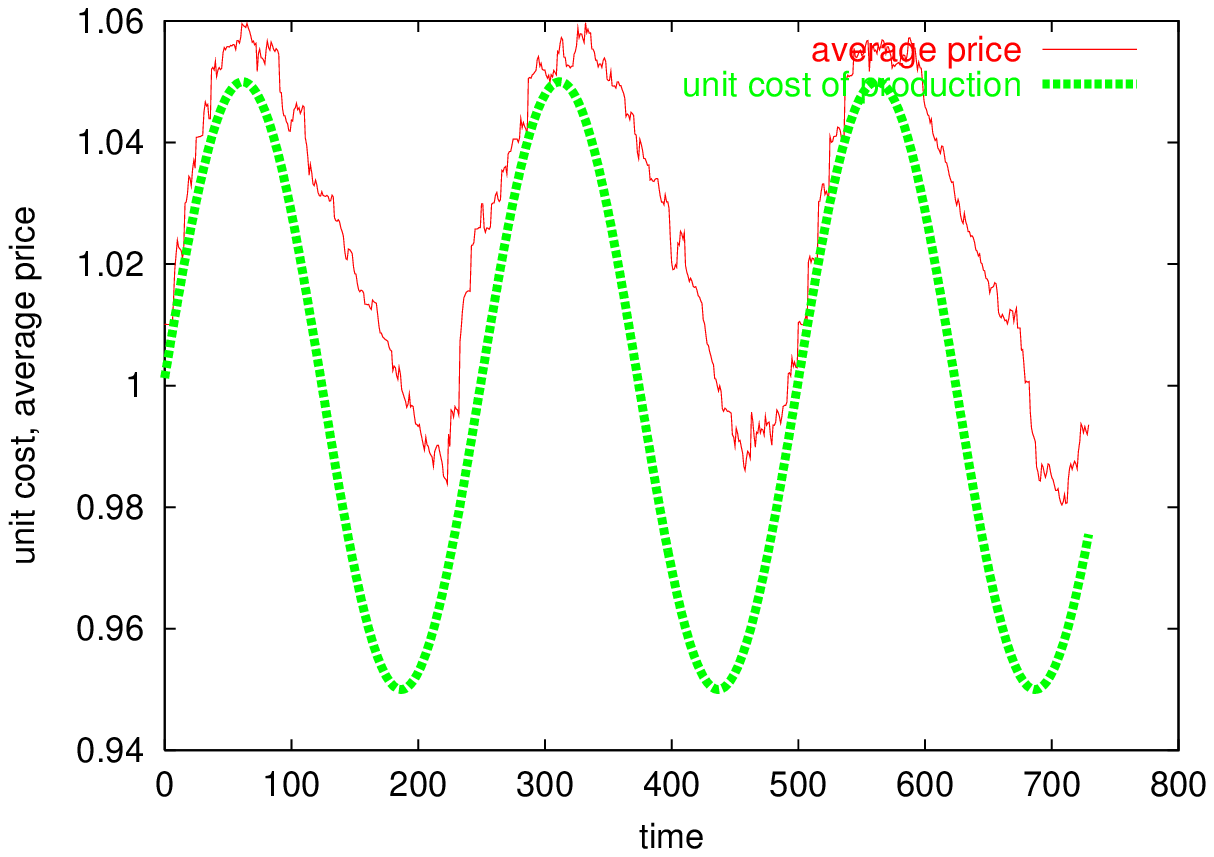}
}
    \caption{%
LEFT: Price adjustment.  Bottom curve: when searching consumers compare
      prices.  Top curve: when searching consumers accept
      prices no matter what they are.  RIGHT: Prices tracking the cost of production.
}
    \label{fig:adjustment-1}
  \end{center}
\end{figure}

In the above model, price converges to the unit cost of production,
which is the competetive price.  In Fig.~\ref{fig:adjustment-1} (left,
bottom curve) we show how an initially higher price slowly decreases
towards a price of one.  The reason for this is that, as long as
prices are larger than one, there will be companies that, via random
changes or injection, have a lower price than their neighbors.
Eventually, these neighbors raise prices, thus driving their customers
away and to the companies with lower prices.  If, however, a company
lowers its price below one, then it will
immediately exit after it has attracted at least one customer.\footnote{%
  If \emph{all} prices in the system are more than $\delta$ below one,
  then the model is not well-defined.  In the limit of large systems
% need this because for small systems the only active cluster can
 % decrease prices without injection of others.
  and when starting with prices above one, such a state cannot be
  reached via the dynamics. -- Also note that if the model allowed
  credit, the exit of such a company would be delayed, allowing losses
  for limited periods of time.

}

%%%%%%%%%%%%%%%%%%%%%%%%%%%%%%%%%%%%%%%%%%%%

%%%%%%%%%%%%%%%%%%%%%%%%%%%%%%%%%%%%%%%%%%%%

As already mentioned above, it turns out that the price comparison by
the consumers is not needed at all.  We can replace the rule ``if
price goes up, try to find a better price'' by ``if price goes up, go
to a different shop no matter what the price there''.  In both cases,
we find the alternative shop via our neighbors, as we have done
throughout this paper.  The top curve in Fig.~\ref{fig:adjustment-1}
shows the resulting price adjustment.  Clearly, the price still moves
towards the critical value of one, but it moves more slowly and the
trajectory displays more fluctuations.  This is what one would expect,
and we think it is typical for the situation: If we reduce the amount
of ``rationality'', we get slower convergence and larger fluctuations.

In terms of cluster size distribution, the price model is similar to
the earlier spatial competition model with random injection.  They
would become the same if we separated bankruptcy and price changes.

In Fig.~\ref{fig:adjustment-1}~(right) we also show that our model is
able to track slowly varying costs of production.  For this, we
replace $C(Q) = Q$ by a sinus-function which oscillates around $Q$.
The plot implies that prices lag behind the costs of production.

%%%%%%%%%%%%%%%%%%%%%%%%%%%%%%%%%%%%%%%%%%%%

\begin{figure}[t]
  \begin{center}
\centerline{
    \includegraphics[width=0.49\hsize]{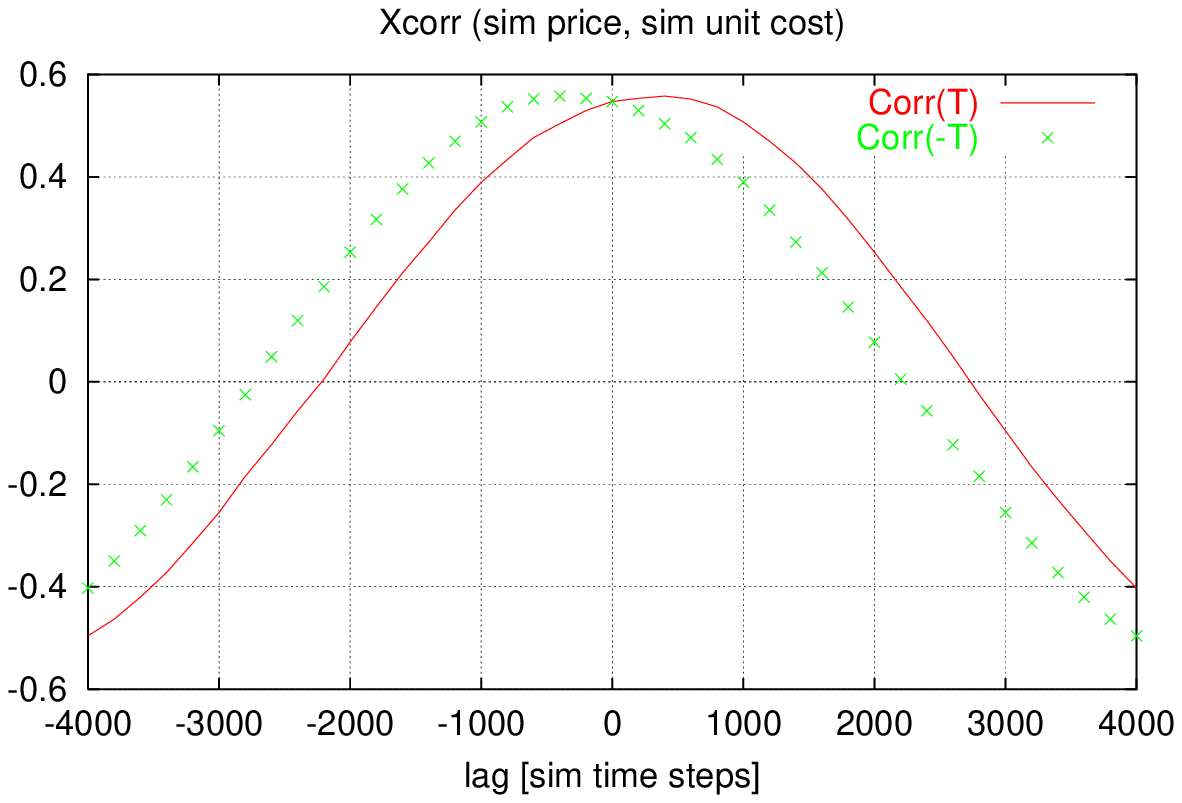}
\hfill
    \includegraphics[width=0.49\hsize]{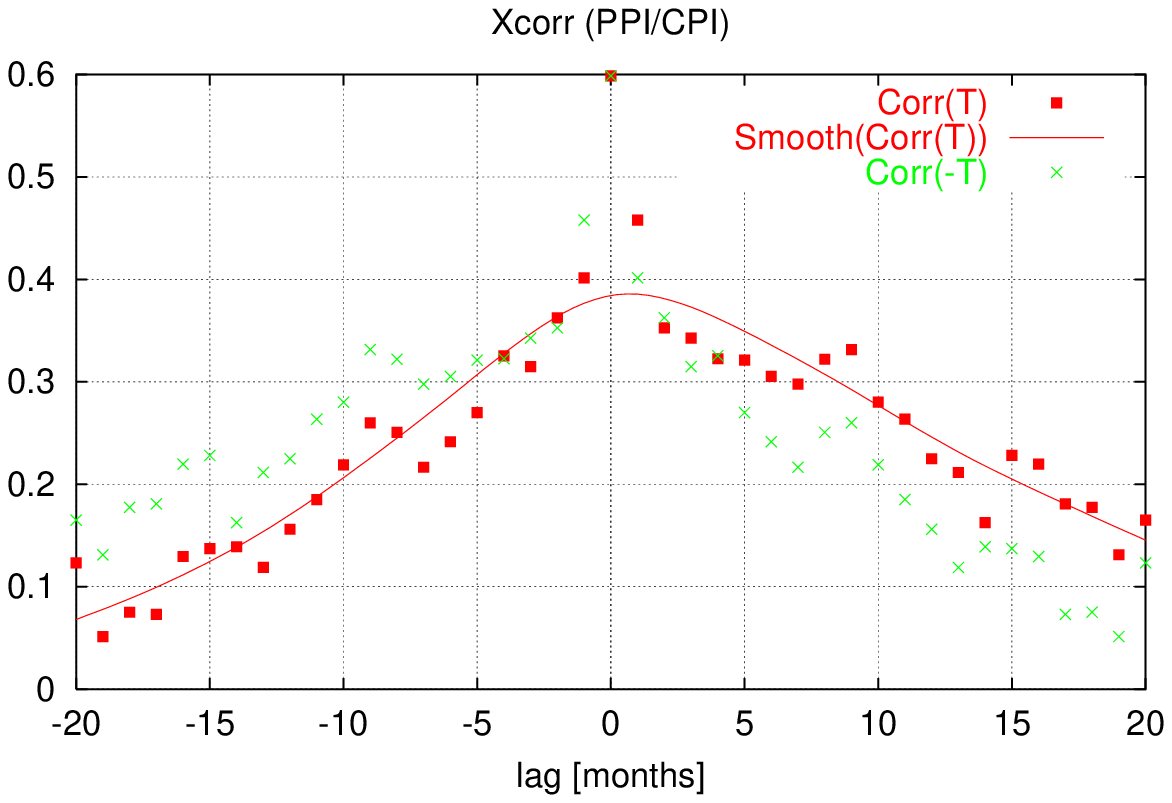}
}
    \caption{%
      Crosscorrelation function between $R_{\it price}$ and $R_{\it
        cost}$: $R_x := x(t) / x(t\!-\!1)$;
      ${\it Xcorr}(\tau) := \langle R_P(t) R_C(t\!-\!\tau) \rangle$. LEFT:
      Simulation.  The crosses show the crosscorrelation values
      mirrored at the $\tau=0$ axis, in order to stress the asymmetry.
      RIGHT: U.S.\ Consumer price index for price and Producer price
      index for cost.  Filled boxes are the crosscorrelation values;
      the smooth line is an interpolating spline for the filled boxes.
      The crosses show the crosscorrelation values mirrored at the
      $\tau=0$ axis.
}
    \label{fig:xcorr}
  \end{center}
\end{figure}

%%%%%%%%%%%%%%%%%%%%%%%%%%%%%%%%%%%%%%%%%%%%

%\yyyy{The rest until the end of the section is in my view a candidate
%  for replacement by a brief summary.  Including the fig this will
%  save about 1/2 page.  Any opinions?  Kai}

This is also visible in the asymmetry of the cross correlation
function between both series.  In order to be able to compare with
non-stationary real world series, we look at relative changes, $R_x(t)
= x(t) / x(t-1)$.  The cross correlation function between price
increases and cost increases then is
\[
{\it Xcorr}(\tau) = \langle R_P(t) \, R_C(t-\tau) \rangle \ ,
\]
where $\langle . \rangle$ averages over all $t$.  In
Fig.~\ref{fig:xcorr}~(left) one can clearly see that prices are indeed
lagging behind costs for our simulations.  In order to stress the
asymmetry, we also plot ${\it XCorr}(-\tau)$.  In
Fig.~\ref{fig:xcorr}~(right) we show the same analysis for the
Consumer Price Index vs.\ the Production Price Index (seasonally
adjusted).  Although the data is much more noisy, it is also clearly
asymmetric.

%%%%%%%%%%%%%%%%%%%%%%%%%%%%%%%%%%%%%%%%%%%%

%%  \begin{figure}[t]
%%    \begin{center}
%%      \caption{}
%%      \label{fig:track}
%%    \end{center}
%%  \end{figure}

%%  \begin{figure}[p]
%%    \begin{center}
%%      \ncludegraphics[width=0.8\hsize]{field-gz.eps}
%%      \caption{Consumer Price Index, Producer Price Index}
%%      \label{fig:field}
%%    \end{center}
%%  \end{figure}

%%  \begin{figure}[p]
%%    \begin{center}
%%      \includegraphics[width=0.8\hsize]{ch-distrib-field-gz.eps}
%%      \caption{Distribution of $C(t)/C(t-1)-1$ (red) and $P(t)/P(t-1)-1$ (green).}
%%      \label{fig:ch-field}
%%    \end{center}
%%  \end{figure}

\section{Discussion and outlook}
\label{sec:discussion}

The modelling approach with respect to economics in this paper is
admittedly simplistic.  Some obvious and necessary improvements
concern credit and bankruptcy (i.e.\ rules for companies to operate
with a negative amount of cash).  Instead of those, we want to discuss 
some issues here that are closer to this paper.  These issues are
concerned with time, space, and communication.  

In this paper, in order to reach a clean model with possible analytic
solutions, we have described the models in a language which is rather
unnatural with respect to economics.  For example, instead of ``one
company per time step'' which changes prices one would use rates (for
example a probability of $p_{\it ch}$ for each company to change prices in
a given time step).  However, in the limiting case of $p_{\it ch} \to 0$,
at most one and usually zero companies change prices in a given time
step.  If one also assumes that consumers adaptation is fast enough so 
that it is always completed before the next price change occurs, then
this will result in the same dynamics as our model.  Thus, our model
is not ``different'' from reality, but it is a limiting case for the
limit of fast customer adaptation and slow company adaptation.  Our
approach is to understand these limiting cases first before we move to 
the more general cases.

Similar comments refer to the utilization of space.  We have already
seen that moving from a spatial to a non-spatial model is rather
straightforward.  There is an even more systematic way to make this
transition, which is the increase of the dimensions.  In two
dimensions on a square grid, every agent has four nearest neighbors.
In three dimensions, there are six nearest neighbors.  In general, if
$D$ is the dimension, there are $2D$ nearest neighbors.  If we leave
the number $N$ of agents constant and keep periodic boundary
conditions ($D$-dimensional torus), then at $D=(N-1)/2$ everybody is a
nearest neighbor of everybody.  Thus, a non-spatial model is the $D
\to \infty$ limiting case of a spatial model.\footnote{%
Furthermore, models such as the ones discussed in this paper often
have a so-called upper critical dimension, where some aspects of the
model become the same as in infinite dimensions.  This upper critical
dimension often is rather low (below 10). 
}

These concepts can be generalized beyond grids and nearest neighbors
-- the only two ingredients one needs is that (i)~the probability to
interact with someone else decreases fast enough with distance, and
that (ii)~if one doubles distance from $r$ to $2r$, then the number of
interactions up to $2r$ is $2^D$ times the number of interactions up
to distance $r$.

This should also make clear that space can be seen in a generalized
way if one replaces distance by generalized cost.  For example, how
many more people can you call for ``20~cents a minute or less'' 
than for ``10~cents a minute or less''?  If the answer
to this is ``two times as many'', then for the purposes of this
discussion you live in a one-dimensional world. 

Given this, it is important to note that we have used space only for
the communication structure, i.e.\ the way consumers aquire
information (by asking neighbors).  This is a rather weak influence of
space, as opposed to, for example, transportation
costs\cite{Hotelling}; it however also assumes a not very
sophisticated information structure, as for example contrast to
today's internet.  The details of this need to be left to future work.

Last, one needs to consider which part of the economy one wants to
model.  For example, a stockmarket is a centralized institution, and
space plays a weak role at best.  In contrast, we had the retail
market in mind when we developed the models of this paper.  In fact,
we implicitely assume perishable goods, since agents have no memory of
what they bought and consumed the day before.  Also, we assume that
consumers spend little effort in selecting the ``right'' place
to shop, which excludes major personal investments such as cars or
furniture.  Also note that our companies have no fixed costs, which
implies that there are no capital investments, which excludes for
example most manufacturing.

\section{Summary}

Price formation is an important aspect of economic activity.  Our
interest was in price formation in ``everyday'' situations, such as
for retail prices.  For this, we assumed that companies are price
setters and agents are price takers, in the sense that their only
strategy option is to go someplace else.  In our abstracted situation,
this means that companies with too low prices will exit because they
cannot cover costs, while companies with too high prices will exit
because they lose their customers.

We use space in order to simplify and structure the way in which
information about alternative shopping places is found.  This prevents
the singularity of ``Bertrand-style'' models, where the market share
of each company is independent from history, leading to potentially
huge and unrealistic fluctuations.

By doing this, one notices that the spatial dynamics can be separated
from the price formation dynamics itself.  This makes intuitively
sense since, in generalized terms, we are dealing with evolutionary
dynamics, which often does not depend on the details of the particular 
fitness function.  We have therefore started with an investigation of
a spatial competition model without prices.  For this model, we have
looked at cluster size distributions, and compared them with real
world company size distributions.  In contrast to investigations in
the literature, which find log-normal distributions, we find a scaling 
law a better fit of our data.  In the models, we find log-normal
distributions or scaling laws, depending on the particular rules.

We then added price formation to our spatial model.  We showed that
the price, in simple scenarios, converges towards the competitive
price (which is here the unit cost of production), and that it is able
to track slowly varying production costs, as it should.  This predicts
that prices should lag behind costs of production.  We indeed find
this in the data of consumer price index vs.\ production price index
for the United States since 1941.

\section*{Acknowledgments}

KN thanks Niels Bohr Institute for hospitality during the summer 1999,
where this work was started.  All of the authors thank Santa Fe
Institute, where some of the authors met, which provided a platform
for continuous discussion, and where some of the work was done in
spring 2000.  We also thank H.\ Flyvbjerg and K.~Sneppen for
invaluable hints and discussions.

\appendix

\section{Converting the aggregated census data}

\paragraph*{Non-equidistant bins}

The size data in the 1992 U.S.\ economic census comes in
non-equidistant bins.  For example, we obtain the number of
establishments with annual sales above 25\,000~k\$, between
10\,000~k\$ and 25\,000~k\$, etc.  For an accumulated function, such
as Fig.~\ref{fig:sales}~(right), this is straightforward to use.  For
distributions, such as Fig.~\ref{fig:sales}~(left), this needs to be
normalized.  We have done this in the following way:
(1) We first divide by the weight of each bin, which is its width.
  In the above example, we would divide by
  $(25\,000~k\$ - 10\,000~k\$) = 15\,000~k\$$.  Note that this
  immediately implies that we cannot use the data for the largest
  companies since we do not know where that bin ends.
(2) For the log-normal distribution
\[
\rho(x) 
\propto {1 \over x} \, \exp\big[ - ( \ln(x) - \ln(\mu) )^2 \big]
\]
(note the factor $1/x$), one typically uses logarithmic bins, since
then the factor $1/x$ cancels out.  This corresponds to a weight of
$x$ of each census data point.
(3) Now we have to decide where we plot the data for a specific
  bin.  We used the arithmic mean between the lower and the upper
  end.  In our example case, $17\,500k\$$.  
(4) In summary, say the number of establishments between $s_i$
  and $s_{i+1}$ is $N_i$.  Then the transformed number $\tilde N_i$ is
  calculated according to
\[
\tilde N_i = {N_i \over s_{i+1} - s_i} \, {s_i + s_{i+1} \over 2} \ .
\]

\paragraph*{The largest firms}

For the largest firms (but not for the large establishments), the
census also gives the combined sales of the four (eight, twenty,
fifty) largest firms.  We used the combined sales of the four largest
firms divided by four as a (bad) proxy for the sales of each of these
four companies.  We then substracted the sales of the four largest
firms from the sales of the eight largest firms, divided again, etc.
Those data points should thus be seen as an indication only, and it
probably explains the ``kink'' near $2 \times 10^9$ in
Fig.~\ref{fig:sales}.

%***********************************************************************
\bibliographystyle{unsrt}
\bibliography{ref,kai,xref}

\begin{thebibliography}{10}

\bibitem{Hotelling}
H.~Hotelling.
\newblock Stability in competition.
\newblock {\em Economic Journal}, 39(41):57, 1929.

\bibitem{Kiyotaki:Wright:money}
N.~Kiyotaki and R.~Wright.
\newblock On money as a medium of exchange.
\newblock {\em Journal of Political Economy}, 97(927):934, 1989.

\bibitem{Bak:etc:money}
P.~Bak, S.F. N\/orrelykke, and M.~Shubik.
\newblock Dynamics of money.
\newblock {\em Physical Review E}, 60(3):2528--2532, 1999.

\bibitem{Sneppen:money}
R.~Donangelo and K.~Sneppen.
\newblock Self-organization of value and demand.
\newblock cond-mat preprint 9906298, arXiv.org, 1999.

\bibitem{Cournot:book}
A.A. Cournot.
\newblock {\em Researches into the Mathematical Principles of the Theory of
  Wealth (translated from French; original: 1838)}.
\newblock Macmillan, New York, 1897.

\bibitem{Bertrand}
J.~Bertrand.
\newblock Theorie mathematique de la richesse sociale (review).
\newblock {\em Journal des Savants (Paris)}, 68(499):508, 1883.

\bibitem{Chamberlin:book}
E.H. Chamberlin.
\newblock {\em Theory of monopolistic competition}.
\newblock Harvard University Press, Cambridge, MA, 1933.

\bibitem{Hehenkamp:etc:note}
B.~Hehenkamp and W.~Leininger.
\newblock A note on evolutionary stability of {B}ertrand equilibrium.
\newblock {\em Journal of Evolutionary Economics}, 9:367--371, 1999.

\bibitem{Hehenkamp:sluggish}
B.~Hehenkamp.
\newblock Sluggish consumers: {A}n evolutionary solution to the {B}ertrand
  paradox.
\newblock Technical Report 99-04, Microeconomics, University of Dortmund,
  Germany, 1999.

\bibitem{Brenner:prices}
Th. Brenner.
\newblock The dynamics of prices -- {C}omparing behavioural learning and
  subgame perfect equilibrium.
\newblock Papers on Economics and Evolution 0001, Max-Planck-Inst. for Economic
  Systems, Jena, Germany, 2000.

\bibitem{Flyvbjerg:foams}
H.~Flyvbjerg.
\newblock Model for coarsening froths and foams.
\newblock {\em Physical Review E}, 47(6):4037--4054, 1993.

\bibitem{econ:census:92}
U.S.~Census Bureau.
\newblock 1992 {C}ensus of retail trade, volume~2, subject series, part~3.
\newblock www.census.gov/prod/1/bus/retail/92subj/rc92s01.pdf.

\bibitem{Stanley:sizes:econ-letters}
M.H.R. {Stanley et al}.
\newblock Zipf plots and the size distribution of firms.
\newblock {\em Economics Letters}, 49:453--457, 1995.

\bibitem{Sneppen:etc:fractal-cracks}
G.~Huber, M.~Jensen, and K.~Sneppen.
\newblock A dimension formula for self-similar and self-affine fractals.
\newblock {\em Fractals}, 3(3):525--531, 1995.

\end{thebibliography}
%***********************************************************************

\end{document}